\documentclass[10pt,conference]{IEEEtran}
\newtheorem{definition}{Definition}

\usepackage{amsmath}
\usepackage{array}

\usepackage{graphicx}
\usepackage[caption=false,font=footnotesize]{subfig}

\usepackage{dblfloatfix}

\usepackage{url}

\usepackage{cite}

\begin{document}

\title{Contrasting Web Robot and Human Behaviors \\with Network Models}

\author{\IEEEauthorblockN{Kyle Brown}
\IEEEauthorblockA{Department of Computer Science and Engineering \\ Kno.e.sis Research Center\\
	Wright State University, Dayton, OH, USA\\
Email: \texttt{brown.718@wright.edu}}
\and
\IEEEauthorblockN{Derek Doran}
\IEEEauthorblockA{Department of Computer Science and Engineering \\ Kno.e.sis Research Center\\
	Wright State University, Dayton, OH, USA\\
Email: \texttt{derek.doran@wright.edu}}}

\maketitle

\begin{abstract}
The web graph is a commonly-used network representation of the hyperlink structure 
of a website. A network of similar structure to the web graph, which we 
call the \emph{session graph} has 
properties that reflect the browsing habits of the agents 
in the web server logs. In this paper, we apply
session graphs to compare the activity of humans
against web 
robots or crawlers. 
Understanding these properties will enable us to improve models of HTTP traffic, 
which can be used to predict and generate realistic traffic
for testing and improving web server efficiency, as well as devising new caching algorithms. 
We apply large-scale network properties, such as the connectivity and degree distribution of human
and Web robot session graphs in order to identify characteristics of the traffic
which would be useful for modeling web traffic and improving cache performance. We find that
the empirical degree distributions of session graphs for human and robot requests on one 
Web server are best fit by different theoretical distributions, indicating at a difference
in the processes which generate the traffic.
\end{abstract}

\section{Introduction}
\label{sec:introduction}
\IEEEPARstart{A}{gents} accessing the World Wide Web (WWW) can be placed 
into two broad categories - human and robot. A robot is defined as  
software that makes requests to a web server without direct human
interaction. Examples of robots are web {\em indexers}, which are employed by 
search engines to build their indexes; 
{\em scrapers}, which can mass download
large numbers of pages for offline storage;
and {\em link checkers}, which verify
that hyperlinks on a web page are still valid over time. There is evidence that traffic from web
robots will only increase over time. 
A 2004 study found that Web robots
accounted for 8.51\% of HTTP requests and 0.65\% of bytes transferred, while
in a 2008 study they made up 18.5\% of HTTP requests and 3.52\% of bytes 
transferred~\cite{doran2011web}. More recently, studies have found Web
robots to make up over 50\% of traffic in different 
domains~\cite{doran2013comparison}. A growing source of web robot traffic
comes from Internet of Things (IoT) devices, which can send large amounts
of traffic from real-time sensors~\cite{chen2012challenges}. Another source are mobile devices, whose traffic volume is growing faster
than traditional broadband~\cite{maier2010first}. 

Recognizing that most web server system optimizations expect
traffic that is statistically and behaviorally 
human-like, it is important to recognize the 
similarities and differences between human and robot
traffic. 
For example, a common technique used by web servers improve response times
is caching, which allows commonly requested
resources to be served faster by storing them in some limited but fast memory. 
Since cache space is limited, a standard approach will not be able to keep up 
with traffic that requests a wide variety of resources unpredictably. Previous
studies suggest that web robot traffic can have a negative impact on
cache performance~\cite{doran2013comparison}. This means that traditional 
approaches to caching will perform worse as the proportion of web robot traffic
increases.

In order to mitigate cache performance degradation, the characteristics of
web robot traffic which set it apart from human traffic must be understood so
that new caching policies and algorithms can be devised which are able to 
handle traffic from robots. Knowledge of the characteristics of web robots
could also be used to create models to simulate web traffic, which could be
used to test new caching algorithms without requiring real traffic. Traffic
simulation would also give the cache algorithm designers control over the
properties of the generated traffic, allowing them to test certain aspects
of their algorithms under known conditions.

A common model of the World Wide Web (WWW) is as a directed network known as
the \emph{web graph}~\cite{broder2000graph,kleinberg1999web,kumar2000web}. 
The nodes of the network are resources such as images,
HTML pages, JavaScript resources, etc. An edge is drawn from an HTML resource
to any other resource if there is a hyperlink from that HTML resource to the
other resource. Under the assumption that agents follow the link structure
of a website, it may be able to reconstruct portions of the network structure
of a website from raw server log data. Building this network using
a notion of sessions is the starting point for our work.

This paper presents an analysis of the characteristics of web robot and human
traffic using metrics from network science. Most of the measures assign a
number to each node, giving an empirical probability distribution over nodes
which can be compared across graphs. We place 
particular focus on modeling the degree distributions
of session graphs as computed from Web server logs. 
Our study confirms the presence of separate mechanisms 
driving the different traffic classes, suggesting that
new forms of web server optimization are needed to handle
traffic that is dominated by robots. 

The rest of this paper is structured as
follows: In the Related Work section, 
references to publications that study the characteristics of the web graph 
and web traffic are given. The Methodology
section introduces the dataset, how it is processed, and the
network metrics computed to compare the two networks. The 
Analysis section describes our results and 
what they tell us about the differences between human and robot requests.
We end with a Conclusion and a discussion of
possible further work in studying session graphs.

\section{Related work}
\label{sec:related_work}
This paper combines two types of studies: characterization of web robot
and human traffic, and analysis of the web graph using network science. 
There have been many studies done on the characteristics of human and web
robot traffic. A classic study on web crawlers is by Dikaiakos \textit{et al.}
~\cite{dikaiakos2005investigation}. The authors use web-server access logs from
academic sites in three countries and compare crawler traffic to general
WWW traffic and devise metrics to provide qualitative characteristics of
crawler behavior. Lee \textit{et al.} published a characterization study of
web robots based on over a billion requests made to \url{microsoft.com}~\cite{lee2009classification}. 
Similar to
~\cite{dikaiakos2005investigation}, they look at the characteristics of specific
web crawlers, and use similar metrics such as response sizes and codes.
In ~\cite{sisodia2015comparative}, Sisodia \textit{et al.} compare the access
behavior of human visitors and web robots through the access logs of a
web portal. They analyze the hourly activity, exit and entry patterns, 
geographic origin of the agents, and the distribution of response
size and response codes by agents. Doran \textit{et al.} studied the 
distributions of response sizes and codes, resource types, and resource
popularities between human and robot agents~\cite{doran2013comparison}.
S. Ihm \textit{et al.} analyzed five years of web traffic 
to determine major changes in the characteristics of this traffic over time~\cite{ihm2011towards}.

Several studies have also been done on properties of the web graph. 
Broder \textit{et al.} analyze the large-scale structure of the web graph, 
showing that it resembles a bowtie with an "IN" component, an "OUT" component,
and a large strongly-connected core~\cite{broder2000graph}. Donato \textit{et al.}
analyze the topological properties of web graphs, including degree distribution, 
PageRank values, and number of connected components~\cite{donato2004large}. A
more recent study confirms the existence of a large strongly connected component,
but indicates that other features such as the "bowtie" structure could be dependent
on the crawling process that produced the web graph~\cite{meusel2014graph}. Sanders and
Kaur use DNS traffic traces to study the graph-theoretic properties of the Web~\cite{sanders2015graph},
in contrast to the more common approach of using HTML pages.
They look at the degree distributions, graph spectrum, clusters, and 
connected components of the resulting web graph.

Liu \emph{et al.} analyze the user browsing graph~\cite{liu2009user}, which is 
similar to the session graph studied in this paper. However, there are two key 
differences: instead of considering sessions they always create an edge for two
requests by the same user in sequence, regardless of the time between the requests.
They also compare their browsing graph to an hyperlink graph, instead of comparing
Web robot browsing graphs to human browsing graphs. They conclude their study by
looking at the PageRank~\cite{page1999pagerank} values of the networks. Computing sessions
and comparing Web robot and human traffic are the novel aspects of our approach.

\section{Methodology}
\label{sec:methodology}
This section introduces the notion of a session graph, the dataset we evaluate robot and
human traffic within, and the metrics considered. First, 
key definitions and concepts which give rise to
the networks we consider are presented. 

\begin{definition}
A \emph{web graph} $G = (V,E)$ is a collection of hyperlinked resources $V$, along
with a set of directed edges $E$, where an ordered pair of resources $(v_1, v_2)$ are
in $E$ if $v_1$ links to $v_2$.
\end{definition}

Note that the web graph is based solely off of the HTML (hypertext) structure of
a website, without any consideration of the agents which visit it.
A session graph is based 
on the identification of user sessions
discussed in~\cite{calzarossa2011analysis} and~\cite{tan2004discovery}. We
give a formal definition of a session below.
\begin{definition}
A \emph{session} $S = (r_1, \dots, r_n)$ of length $n$ is a sequence of resources
$r_i$ requested by the same agent such that if $\tau(r_i)$ is the time at which
resource $r_i$ in the sequence was requested, then for $i=2,\dots,n$, we have
that $\tau(r_i) - \tau(r_{i-1}) < T$ where $T > 0$ is some cutoff time.
\end{definition}
Note that we will often use the word \emph{transition} to mean an ordered pair $(r_i,r_j)$
of resources which appear in sequence within a session. With the concept of a 
session defined, we can now proceed to define a session graph constructed 
from Web server logs.

\begin{definition}
Given a collection of sessions $\mathcal{S}$ and a cutoff time $T > 0$, the
\emph{session graph} defined by $\mathcal{S}$ and $T$ is a tuple $G = (V,E)$
where the vertices $V$ are the resources appearing in the $\mathcal{S}$ and
a directed edge $(r_1, r_2)$ is in $E$ if the sequence $r_1, r_2$ appears
in some session.
\end{definition}

The preceding definitions can be understood more informally as follows. 
A \emph{session} is a sequence of requests made by an agent with the 
same user-agent string (or IP address) such that the time between each 
request is less than some cutoff value $T > 0$.
The nodes of the session graph are all resources appearing in the Web 
server logs. A directed edge is added between two nodes if the
two resources were requested in sequence within a session. 
To identify agents, we use the \texttt{User-Agent} string provided in the
HTTP header along with the IP address.

\subsection{Dataset and Preprocessing}
Our dataset consists of web server access logs from the domain \texttt{wright.edu}
for the months of April, May, and July in 2016. 
A typical entry in the log looks like the following:

\begin{verbatim}
- - [02/Apr/2016:00:00:09 -0400] 
"GET /path/to/some/resource HTTP/1.1" 200  
5972 "http://www.example.com/refererpage.html" 
"Mozilla/5.0 (iPhone; CPU iPhone OS 7_0 
like Mac OS X)" "11.111.111.111"
\end{verbatim}

Each log entry includes at least the time and date the request was observed, 
HTTP request including method, path, and HTTP version, HTTP response
code from the server, and IP address of the requester. \
Other fields which may or may not be present are the response
size, referer field in the HTTP header, and User-Agent string. 
Each file containing the raw server logs is split into two separate files, one containing
only human traffic, and the other containing only robot traffic. This is done using the
crowd-sourced database BotsVsBrowsers~\cite{bvb_site} to identify robots based on the
\texttt{User-Agent} HTTP header field and/or IP address.
We acknowledge that probabilistic methods
exist to better separate robots and humans~\cite{doran2011web};
however, our goal is to extract samples of robot and human 
sessions that are verifiably correct, so such a complicated approach
is not necessary.

Human traffic was extracted from all three months 
of data. Only robot requests for the first 20 days of the month of 
April were used due to computational limitations from 
the large number of robot requests. Since the resulting number of
Web robot requests was still larger than the number of human requests,
and because we don't feel it's likely that the nature of Web robot
traffic would change greatly in 3 months, this does not have a large
impact on our analyses.
Summary statistics of the robot and human traffic are
provided in Table~\ref{tbl:dataset_summary}. Even though more files were
used for humans than robots, there are still more robot requests than human
requests. However, there are less robot sessions. This could indicate that
robots tend to have larger sessions. A similar thing happens with agents and
IP addresses; there are more human agents, but less human IP addresses. This
is probably due to the fact that crawlers tend to have several different IP
addresses, but sharing the same user-agent string. The number of resources is
larger for robots than humans, indicating that robots may tend to request old,
non-existent, or otherwise uncommon resources more often.

\begin{table}[h]
\vspace{-10px}
	\centering
	\caption{Summary of the Dataset}
	\begin{tabular}{|c||c|c|}
	\hline
	\bfseries{Metric} & \bfseries{Humans} & \bfseries{Robots} \\
	\hline
	\# Files & 91 & 20 \\
	\# Requests & 197056 & 427472 \\
	\# Sessions & 23825 & 11259 \\
	\# Agents & 1429 & 330 \\
	\# IP addresses & 2174 & 4211 \\
	\# Resources & 34185 & 75776 \\
	Start time & April 1, 2016 & April 1, 2016 \\
	End time & June 30, 2016 & April 20, 2016 \\
	\hline
	\end{tabular}
	\label{tbl:dataset_summary}
\end{table}

We parsed the Web server logs using a regular expression in Python, then 
used the freely available \texttt{igraph} library~\cite{igraph_site,csardi2006igraph} 
to build the session graph and compute its various properties.

\subsection{Session Graph Metrics}
This section describes the network metrics analyzed and also serves to
clarify the notation used. For an introduction to network science as a whole,
the text by Newman~\cite{newmannetworks} is standard. Other overviews can be
found in~\cite{barabasi2013network,lewis2011network,fang2007new}.

We will denote a directed graph by $G = (V,E)$ where
$V$ is the set of vertices or nodes and $E \subseteq V\times V$ is the set of
directed edges. $n = |V|$ will always be the number of nodes and $m = |E|$ the
number of edges. The principal graph representation used is the adjacency matrix
$A$, which is an $n\times n$ matrix with entries given by
\begin{equation}
A_{ij} =
\begin{cases}
1, & \text{if there is an edge from i to j} \\
0, & \text{otherwise}
\end{cases}
\end{equation}
This work focuses on connectivity measures, which describe
the distribution of edges, their number, and how nodes in a network 
relate to each other. We start with in- and out-degrees, given by
\begin{equation}
k_i^{\text{in}} = \sum_{j=1}^{|V|}A_{ji}
\end{equation}
and
\begin{equation}
k_i^{\text{out}} = \sum_{j=1}^{|V|}A_{ij}
\end{equation}
The in-degree conveys how often a resource was visited after another 
resource within sessions, and the out-degree tells us how many times another
resource was visited after this one. A comparison of the degree distributions
for human and web robot traffic networks can tell us how likely it is they
were generated by the same process, even when the exact nature of the process
is unknown.

Another measure is the \emph{density} of the
network, defined as
\begin{equation}
\rho = \frac{|E|}{|V|(|V| - 1)}
\end{equation}
where the denominator is the total number of possible edges in a directed network
with $|V|$ vertices. This gives an idea of how close to being fully connected the
network is. In terms of the session graph, the density reflects the proportion of
transitions observed out of all possible transitions. For agents  following the
the hyperlink structure of the website, the  graph's density should be
close to that of the underlying Web graph.

One way to define a partition over the vertex set of a network is to consider its
\emph{connected components}. For a directed graph, there are two notions of connectivity;
two nodes $v_i$ and $v_k$ are \emph{weakly connected} if in the graph obtained by
replacing all directed edges with undirected ones, there is a path from $v_i$ to $v_k$.
Then $v_i$ and $v_k$ are \emph{strongly connected} if there is a directed path from
$v_i$ to $v_k$ or if there is a directed path from $v_k$ to $v_i$. Then the weakly (strongly)
connected components of a network $G$ are a set of subgraphs of $G$, $\mathcal{C} = (C_1, \dots, C_k)$
such that the $C_i$ are pairwise disjoint, their union is all of $G$, and such that in each $C_i$
all nodes are weakly (strongly) connected. We investigate the number of connected components and
sizes of connected components of our networks.

For interaction measures, we study
the reflexivity or reciprocity of a directed network, which is given by
\begin{equation}
r = \frac{2\sum_{i=1,j=1}^{|V|}A_{ij}A_{ji}}{|V|(|V|-1)}
\end{equation}
For our networks, this provides a way to measure how often two resources are
requested in order both ways. An example of a reflexive relation in a web graph
would be two HTML pages which link to each other. Reflexive relations in our
network can also appear, for example, when a user clicks a link and then navigates
back to the previous page by either pressing the ``back" button on a web browser or
by clicking a link on the page that leads back.

\section{Comparative Analysis}
\label{sec:analysis}
A summary of various metrics of the networks is presented in Table~\ref{tbl:summary}.
The graph for robots was much larger due to the presence of more robot requests in the
web server logs than human requests. This could also represent the fact that some robots
such as crawlers request resources that are less popular among humans, in order to
crawl as much of the website as possible. Note that even though the network for robots
has more edges than the humans' network, its density is comparable, both being on the
order of $10^{-5}$. The reciprocity for both networks is comparable, and is quite low, indicating
that only 5\% or so of possible reciprocal edges were observed. This means that it is very
unlikely that if two resources are requested in sequence, they will be requested some time
later in the reverse of the original sequence.

\begin{table}[h]
\vspace{-10px}
	\centering
	\caption{Properties of the Graphs}
\resizebox{0.48\textwidth}{!}{
	\begin{tabular}{|c||c|c|c|c|c|}
		\hline
		\bfseries{Network} & \bfseries{\# Nodes} & \bfseries{\# Edges} & \bfseries{Density} &
		\bfseries{Recip.} & \bfseries{$\mathrm{E}$[Degree]}\\
		\hline\hline
		Humans & 93,655  & 118,706 & 1.353e-05 & 0.0532 & 1.2675 \\
		Robots & 179,432 & 377,047 & 1.171e-05 & 0.0511 & 2.1013 \\
		\hline
	\end{tabular}}
	\label{tbl:summary}
\end{table}

The decomposition of a graph into connected components provides a partition on the vertex
set. Since we are working with directed graphs, there are two notions of connectivity, namely 
weak and strong connectedness. We computed the weakly and strongly connected components of the 
networks and analyzed the properties of this decomposition. A summary of measures computed from the
weakly connected components (WCCs) and strongly connected components (SCCs) is provided
in Table~\ref{tbl:connected}.

\begin{table}[h]
\vspace{-10px}
\centering
\caption{Summary of Analysis of Connected Components}
\resizebox{0.48\textwidth}{!}{
\begin{tabular}{|c||c|c|c|c|}
	\hline
	\bfseries{Network} & \bfseries{\# WCCs} & \bfseries{\# SCCs}  & 
	\bfseries{Largest WCC} & \bfseries{Largest SCC} \\
	\hline\hline
	Humans & 3,816 & 19,328  & 83,641 & 74,148 \\
	\hline
	Robots & 1,626 & 8,261 & 177,267 & 171,089 \\
	\hline
\end{tabular}}
\label{tbl:connected}
\end{table}

Notice that despite having much more nodes than the humans' network, the robots' network only
has 1,626 weakly connected components compared to the humans' 3,816. It also has fewer SCCs,
with 8,261 compared to the humans 19,328. This could indicate that robots are more likely to jump from one 
resource to another, even if there are no links, leading to a more connected structure. In both
cases, the largest SCC and largest WCC contains almost all of the nodes of the network. This shows
the existence of a giant connected component, similar to that of the web graph~\cite{broder2000graph,meusel2014graph},
but restricted to a single web server.

\subsection{Community Detection and Visualization}
Due to the large size of the networks, it was not possible to visualize them in their
entirety. The difficulty arises in computing an aesthetically pleasing layout for such 
networks in a reasonable time. This is a well-studied problem
in the mathematical field of graph theory~\cite{diaz2002survey} .
Instead, for each network, the subgraph consisting of the 5000 nodes of highest degree
was selected and the largest connected component of that subgraph was visualized 
in Gephi~\cite{bastian2009gephi} using the ForceAtlas2 algorithm to compute 
the graph layout~\cite{jacomy2014forceatlas2}.

\begin{figure}[h]
		\includegraphics[width=0.4\textwidth]{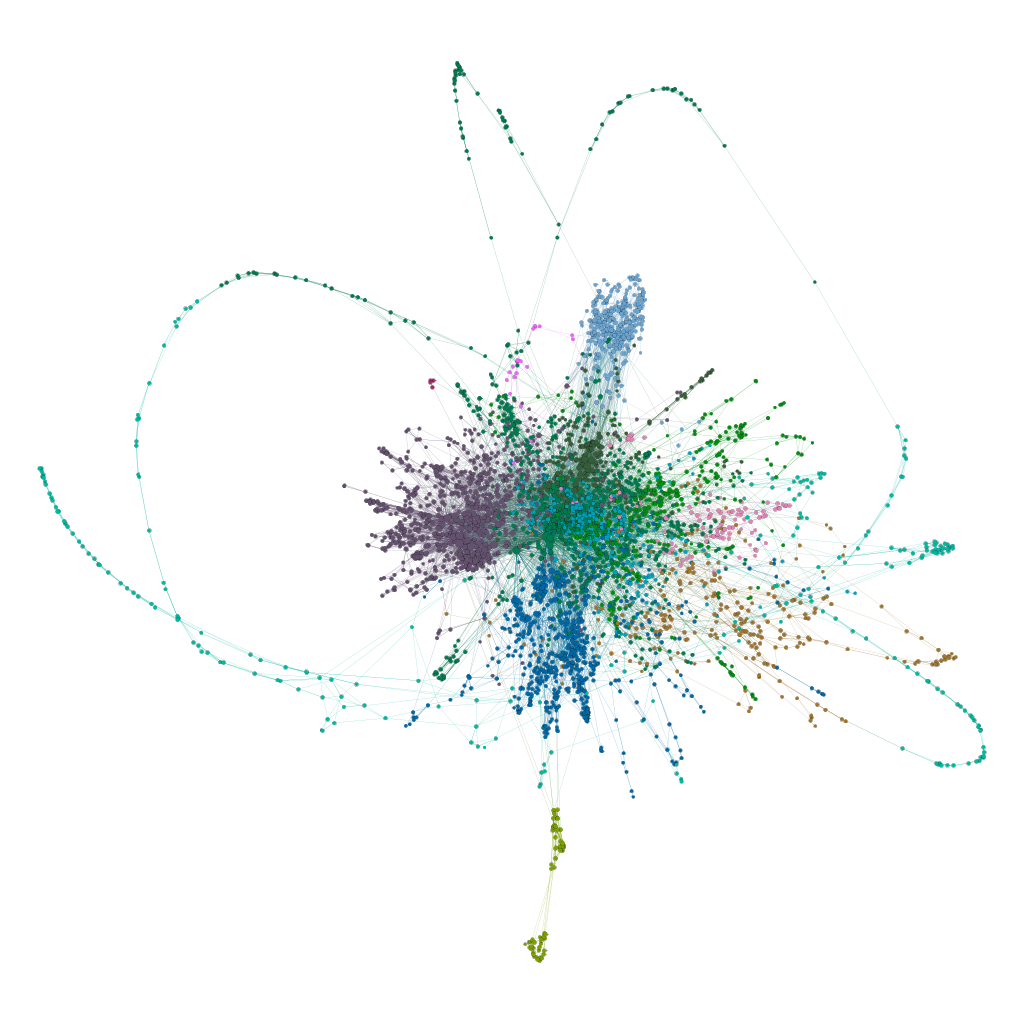}
		\caption{Connectivity among high degree nodes in human session graph}
		\label{fig:hum_vis}
\end{figure}

The humans network is depicted in Figure~\ref{fig:hum_vis}. Nodes are colored by
modularity class, using modularity maximization to compute the community 
structure~\cite{newman2006modularity}. The nodes are sized based on the number of times they were
requested. At the very center of the network is the root node, with path
\texttt{/} on the web server. Much of the resources near the root node are objects
such as JavaScript and CSS files which are automatically loaded by Web browsers
when loading the page.  Since the root node is visited often, these resources end
up having a large number of requests as well. Notice the cluster of beige-colored nodes
which is a little bit more separated from the central agglomeration of clusters. These nodes represent calendar pages which give the dates of important events at the 
university and is used across a number of web pages.
There are also several ``ribbons" of resources which may 
be an artifact of the
process of not visualizing the entire network. These are nodes which are visited in
sequence with nodes of low degree, but which have high degree themselves. When constructing
the subgraph used in visualization, these low degree nodes are left out, isolating the sequences
of high degree resources.

\begin{figure}[h]
		\includegraphics[width=0.4\textwidth]{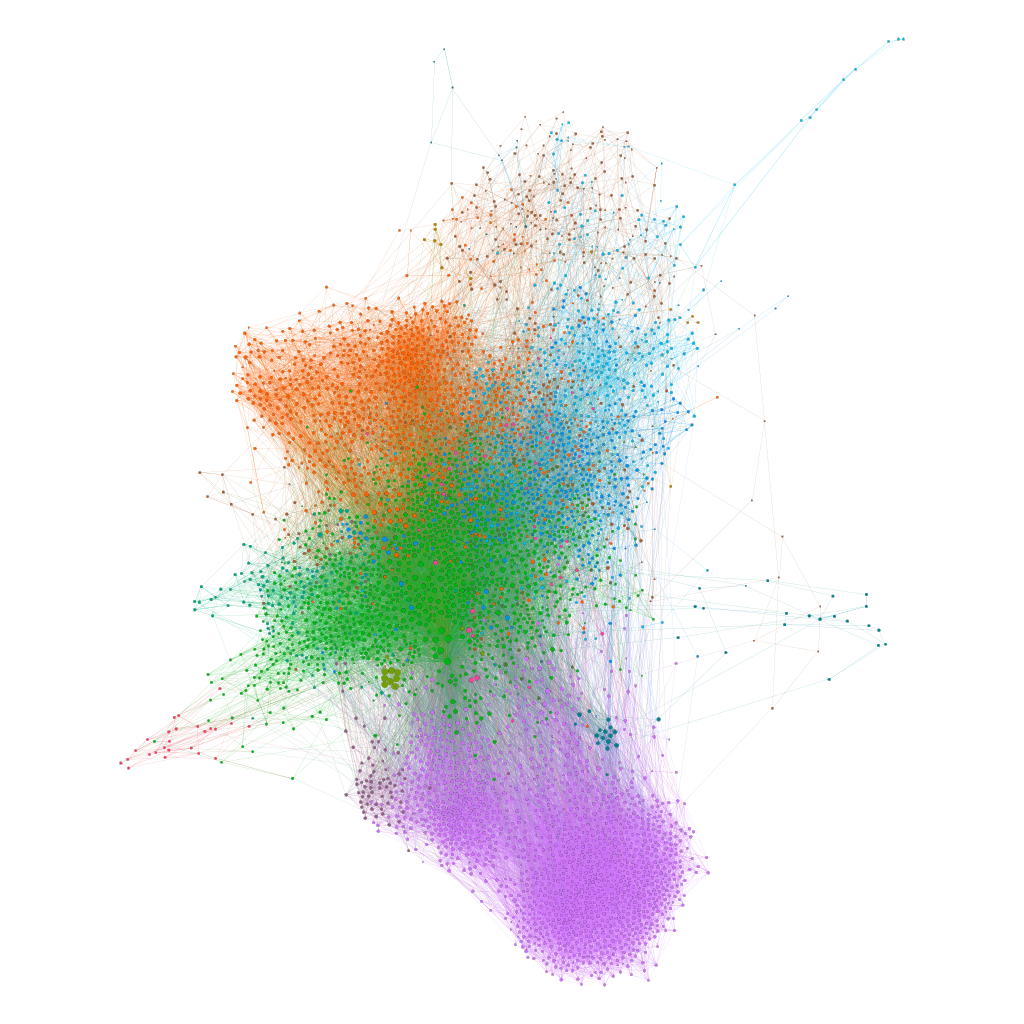}
		\caption{Connectivity among high degree nodes in robot session graph}
		\label{fig:rob_vis}
\end{figure}

The robots network is depicted in Figure~\ref{fig:rob_vis}. Nodes are colored by
modularity class in this visualization as well, and node sizes are based on the
number of requests for the resource. The central green cluster is the ``core" of
the \url{wright.edu} domain, including the \texttt{robots.txt} file. The purple
cluster at the bottom are the contents of personal webpages of faculty and students.
The orange cluster in the middle-upper left comprises the calendar pages, which
are often linked to by news and events on the front page. There are less ribbons
and flares in this visualization, indicating that the highest degree nodes in
the robots network are more interconnected than those in the humans network.

\vspace{-10px}

\subsection{Degree Distributions}
The distribution of degrees in a network can provide much information on how
well-connected the network is, among other things. It has
often been observed
that the 
degree distributions of Internet-related networks tend
to follow power laws~\cite{faloutsos1999power,clauset2009power,adamic2000power}. Power laws are subsets of heavy-tailed
distributions, which is any distribution over a random variable $X$ for which
\begin{equation}
\lim_{x\to +\infty}\mathrm{e}^{\lambda x}\Pr(X \ge x) = \infty
\end{equation}
holds. Heavy-tailed distributions follow a power law
when, for $\alpha > 0$, we have 
\begin{equation}
\Pr(X \ge x) \sim x^{-\alpha}
\end{equation}
Heavy-tailed distributions have a significant non-zero probability 
for large values compared to distributions such as the exponential and 
normal. A key characteristic of power laws is that their
probability mass functions (PMFs) are shaped linearly
when plotted on a loglog scale. In practice it is
difficult to identify power laws, since that many other distributions look nearly linear on a loglog plot. 
Furthermore, the right-tail often exhibits high
variance in many empirical data sets~\cite{clauset2009power}, making it hard to distinguish
between heavy-tail and power-tail behavior. We observe that the 
degree distributions of the networks all exhibit at least heavy-tailed or sub-exponential 
behavior~\cite{foss2011introduction}. 

We compare four candidate distributions to determine which one best matches the
empirical distribution of degrees: exponential, log-normal, Zeta (power law), and double Pareto
log-normal (DPLN) distributions. The Zeta distribution is the discrete analogue of the Pareto 
distribution with parameter $\alpha$, and has PMF
\begin{equation}
f(x;\alpha) = \frac{x^{-\alpha}}{\zeta(\alpha)}
\end{equation}
where $\zeta(\alpha)$ is the Riemann zeta function, defined for  $\alpha > 1$:
\begin{equation}
\zeta(\alpha) = \sum_{n=1}^{\infty}n^{-\alpha}
\end{equation}
The Zeta distribution was chosen as a candidate distribution as it is the simplest discrete
distribution exhibiting power law behavior. When describing discrete, non-negative values 
such as network degrees, a power law is 
preferred over a log-normal because a random variable drawn from the latter can
take on real values, and depending on the parameters, may even have negative values.

The DPLN is a continuous distribution with four parameters, $\alpha, \beta, \mu,$ and $\sigma$, and
has PDF
\begin{align}
f(x) = \frac{\alpha\beta}{\alpha+\beta}\left[
x^{-\alpha-1}\exp\left\{\alpha\mu+\frac{\alpha^2\mu^2}{2}\right\}\right.\nonumber
\mathrm{\Phi}\left(\frac{\log{x} - \mu - \alpha\sigma^2}{\sigma}\right) +\\
\left. x^{\beta-1}\exp\left\{-\beta\mu+\frac{\beta^2\mu^2}{2}\right\}
\mathrm{\Phi}^C\left(\frac{\log{x}-\mu+\beta\sigma^2}{\sigma}\right)
\right]
\end{align}
where $\mathrm{\Phi}$ is the cumulative distribution of the 
standard Normal distribution 
$\mathcal{N}(0,1)$, and $\mathrm{\Phi}^C(x) = 1 - \mathrm{\Phi}(x)$. A derivation of the DPLN, its properties, and some of its applications
can be found in~\cite{reed2004double,zhang2015double}. The DPLN was chosen as a candidate
distribution based on the observation of a noticeable ``bend" in the plots of empirical degree distributions which will
be shown in the sequel. 

\begin{figure*}[h]
	\centering
	\subfloat{
		\includegraphics[width=0.35\textwidth]{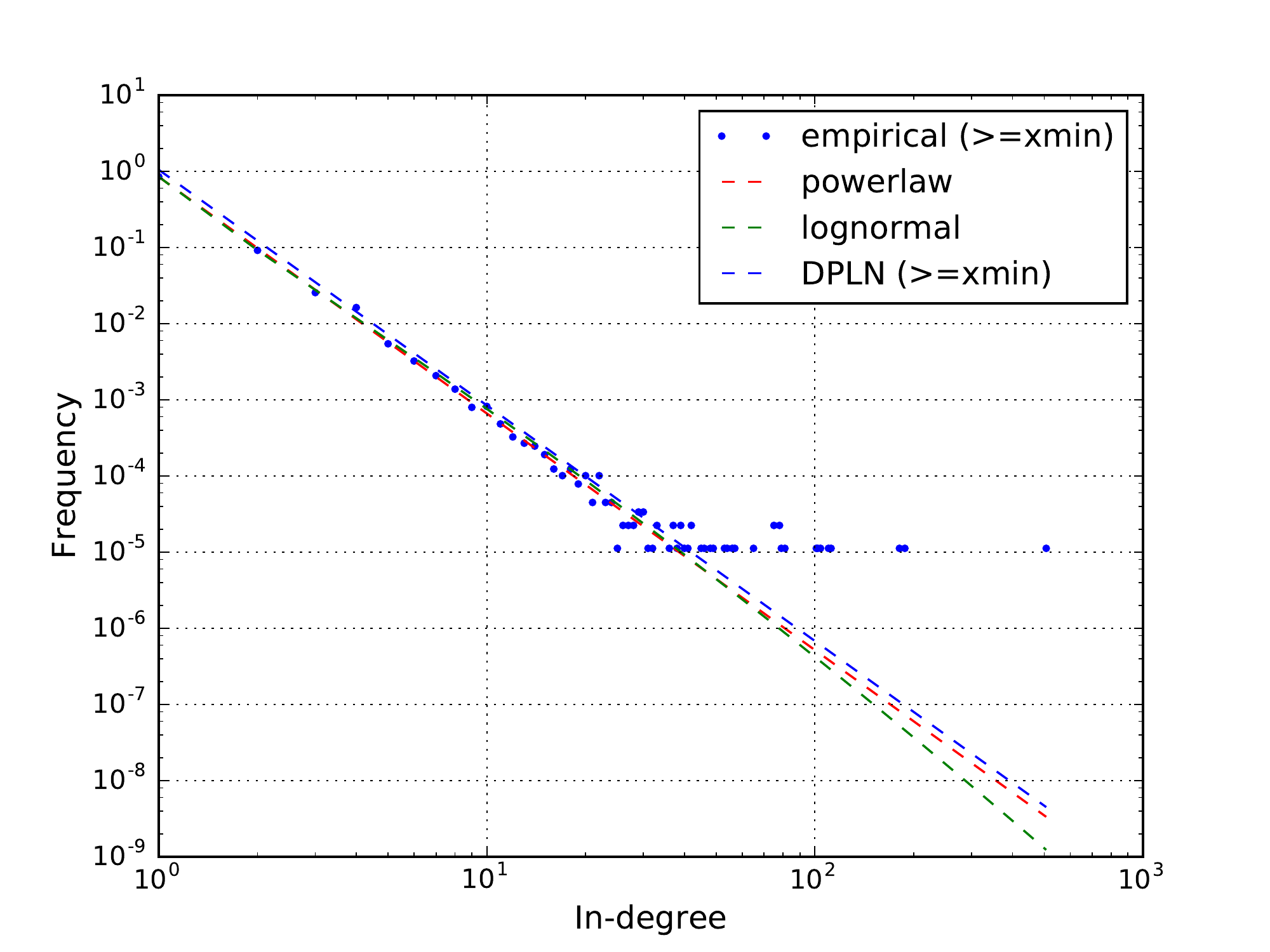}
		\label{fig:hum_indeg}}
	\subfloat{
		\includegraphics[width=0.35\textwidth]{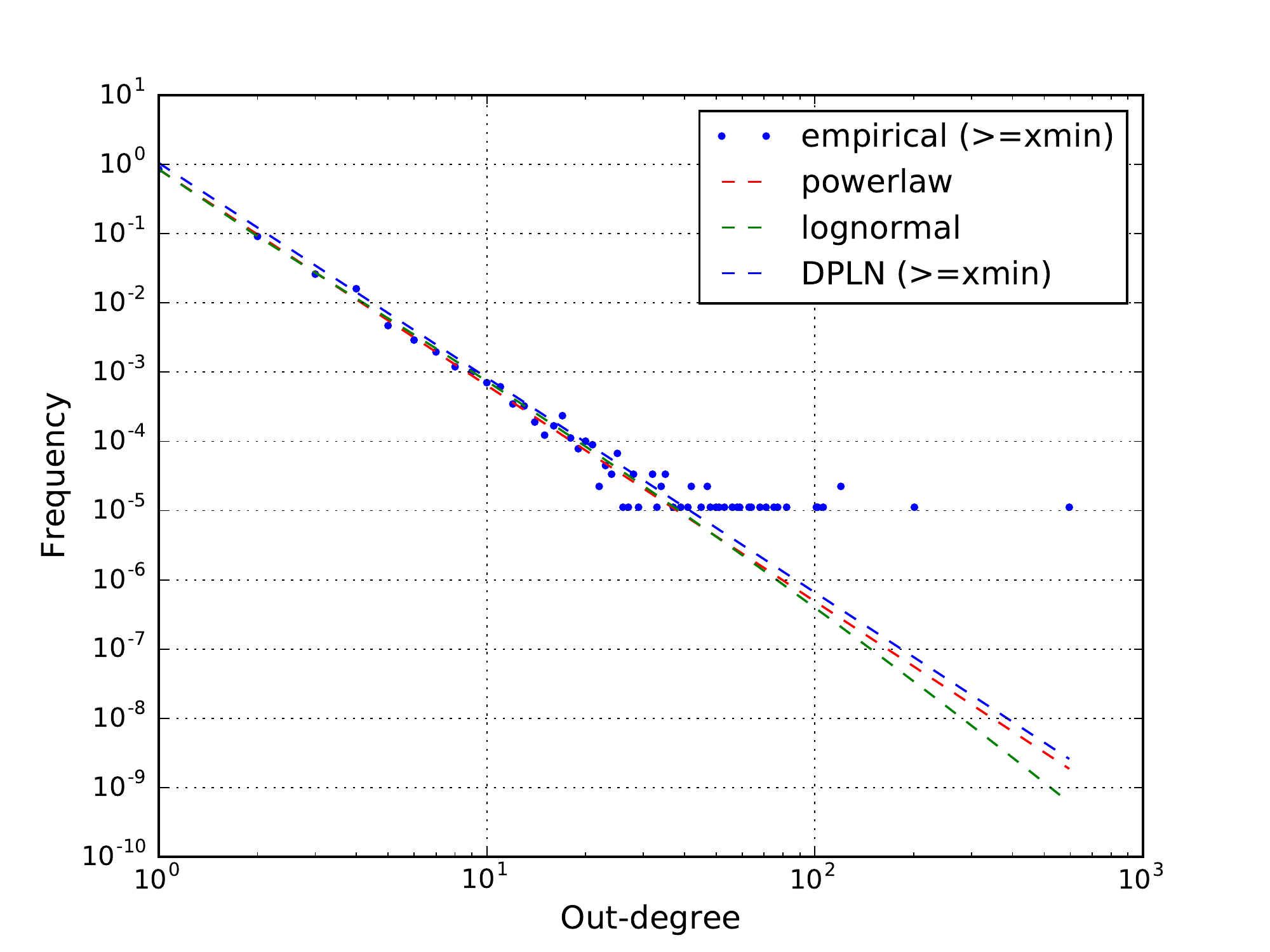}		
		\label{fig:hum_outdeg}}
	\caption{Frequency plots of the humans network's degree distributions}
\end{figure*}

\begin{figure*}[h]
	\centering
	\subfloat{
		\includegraphics[width=0.35\textwidth]{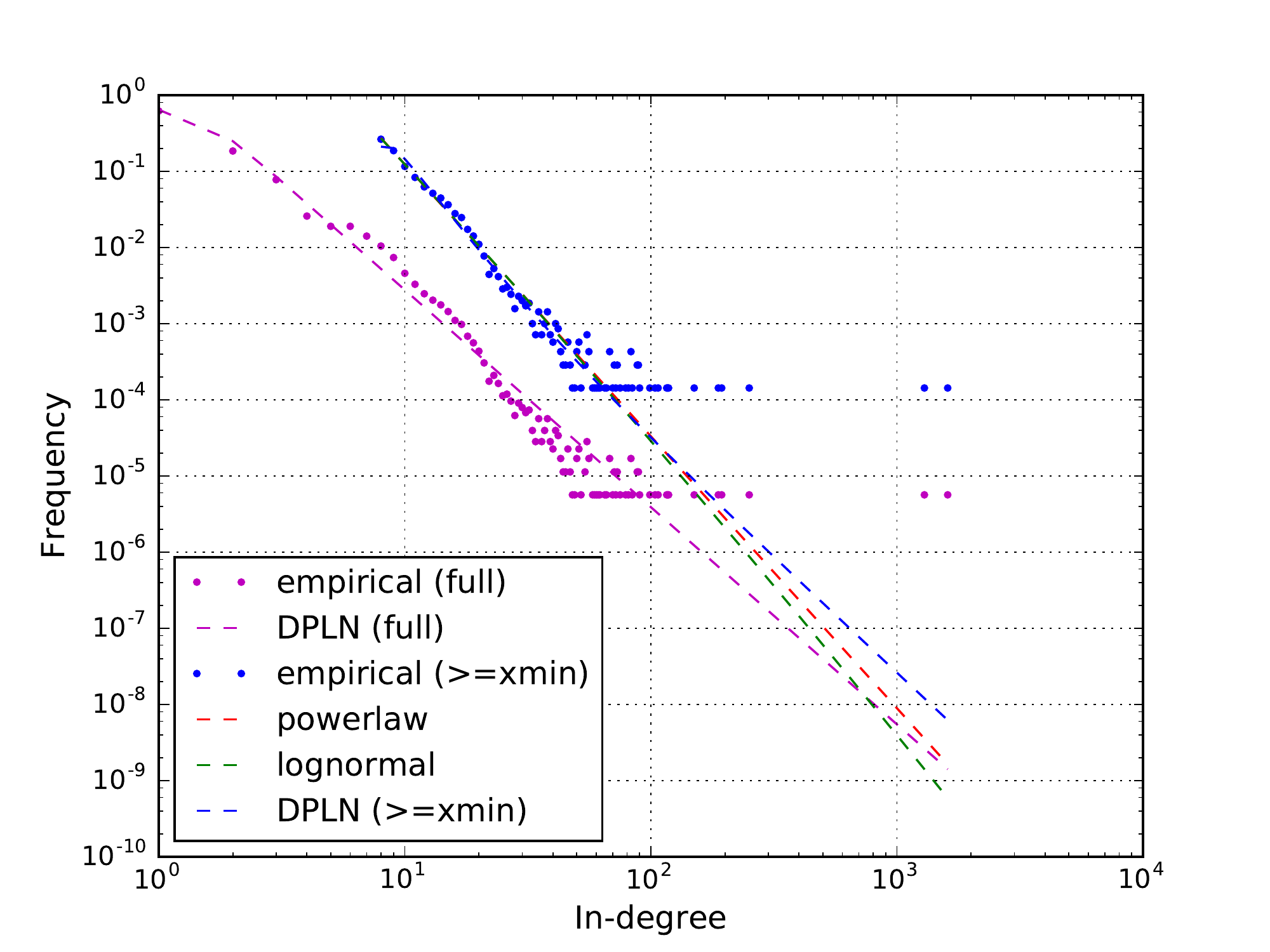}
		\label{fig:rob_indeg}}
	\subfloat{
		\includegraphics[width=0.35\textwidth]{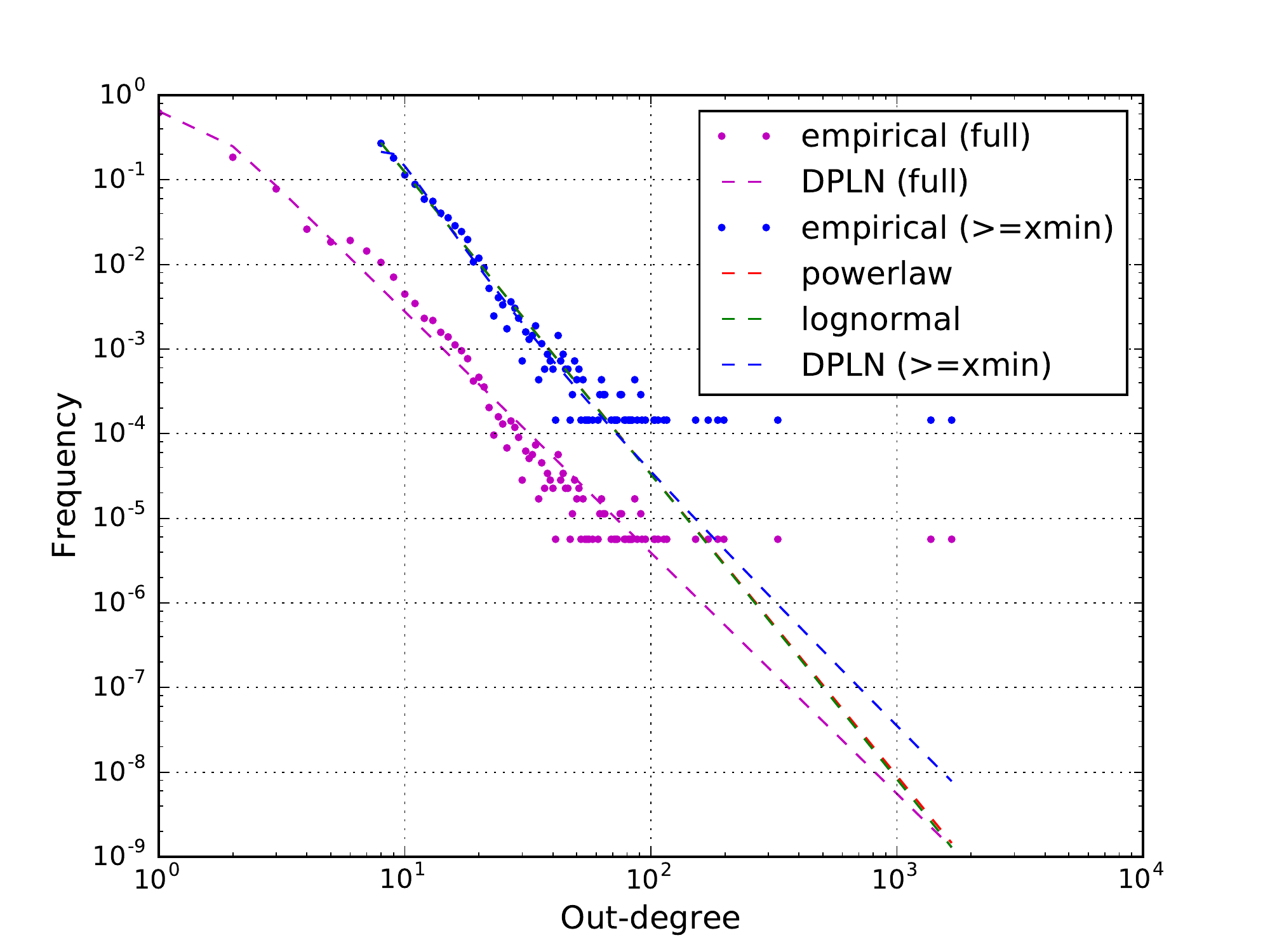}
		\label{fig:rob_outdeg}}
	\caption{Frequency plots of the robots network's degree distributions}
\end{figure*}

Summaries of the maximum likelihood estimates for the log-normal
and Zeta parameters are given in Table~\ref{tbl:dd_summary}.
The degree distributions for the human networks are shown in Figures~\ref{fig:hum_indeg}
and~\ref{fig:hum_outdeg}, and the distributions for robots in Figures~\ref{fig:rob_indeg}
and~\ref{fig:rob_outdeg}. In each plot, a Zeta distribution is fit using maximum likelihood
estimation to the empirical data, and shown alongside it. A log-normal distribution is also
fit, since noise in the right tail can obscure behavior that would distinguish
between a log-normal distribution and a power law~\cite{alstott2014powerlaw}. 
The DPLN is approximated using the method of moments 
as described in ~\cite{reed2004double}. 
The DPLN distribution is notable for exhibiting 
power-tail behavior in both directions, while being similar to the log-normal in 
between~\cite{seshadri2008mobile}. The two parameters $\sigma$ and $\mu$ control the log-normal
behavior, and the parameters $\alpha$ and $\beta$ affect the power-tail (Pareto) behavior.
Power-law-like behavior is most apparent 
in the human degree distribution plots,
which appear nearly linear up to the tail of the empirical distribution. Notice also the 
``bend" in the robot empirical degree distributions, which seems to indicate behavior more
complicated than just a power law.

Note that in the case of the robots plot, the empirical dataset is plotted twice. This is because
the estimated value at which power-tail behavior begins, $x_{min}$, was not 1, so that there were
values below $x_{min}$ which had to be excluded when fitting the power law and log-normal. Therefore,
power law and log-normal distributions are only plotted for degrees greater than or equal to $x_{min}$,
while the DPLN is fitted and plotted for degrees greater than or equal to $x_{min}$ and for all degrees.

\vspace{-10px}
\begin{table}[h]
\centering
\caption{Log-Likelihoods on degree distributions for humans}
\resizebox{0.48\textwidth}{!}{
\begin{tabular}{|c||c|c|c|}
	\hline
	\bfseries{Distributions} & \bfseries{In/out} & \bfseries{$R$} & \bfseries{$p$-value} \\
	\hline
	Exponential-Power law    & in & -10941.4470 & 1.3979e-33 \\
	Lognormal-Power law      & in & 25.7459     & 0.0006     \\
	Lognormal-Exponential    & in & 10967.1929  & 9.7510e-34 \\
	DPLN-Power Law           & in & 19362.2517  & 0.0        \\
	DPLN-Lognormal           & in & 19336.5058  & 0.0        \\
	Exponential-Power law    & out & -10779.5440 & 1.7099e-33 \\
	Lognormal-Power law      & out & 21.1163    & 0.0048 \\
	Lognormal-Exponential    & out & 10800.6603 & 1.2665e-34 \\
	DPLN-Power Law           & out & 19404.8105 & 0.0 \\
	DPLN-Lognormal           & out & 19383.6942 & 0.0 \\
	\hline
\end{tabular}}
\label{tbl:hum_dd_comp}
\end{table}

We evaluated goodness-of-fit for the candidate degree distributions by performing log-likelihood ratio
tests~\cite{vuong1989likelihood} for each pair of candidate distributions. 
The results are shown in Table~\ref{tbl:hum_dd_comp} for
humans and Table~\ref{tbl:rob_dd_comp} for robots. A positive value of $R$ means that the first distribution
is a better fit in the sense of having a larger data likelihood; a negative value of $R$ indicates that the
second distribution is a better fit. The null hypothesis is that the two distributions fit the data equally
well, and the alternative is that one is a better fit than another. We reject the null hypothesis for $p$-values
less than 0.05, which gives $95\%$ significance. Only the observed degrees of value $x_{min}$ or larger were used for
testing, since it would not be possible to compare a DPLN distribution fit to the entire range of degrees
to power laws which are not valid for values less than $x_{min}$.

In all cases, an exponential distribution can be rejected
with high confidence, which is a baseline for showing empirically that a set of samples follows a heavy-tailed
distribution. That is, an exponential distribution being a better fit than any heavy-tailed distribution would
be indicative of non-heavy-tailed behavior.
In the case of humans, a DPLN distribution is the best fit, followed by a log-normal and finally a
power law. With robots, a DPLN distribution can be rejected, but the log-likelihood test could not establish
a significant difference between the log-normal and power law fits. We conclude that log-normal and zeta
distributions describe the robots degree distribution equally well.

\begin{table}[h]
\centering
\caption{Log-Likelihoods on theoretical degree distributions for robots}
\resizebox{0.48\textwidth}{!}{
\begin{tabular}{|c||c|c|c|}
	\hline\bfseries{Distributions} & \bfseries{In/out} & \bfseries{$R$} & \bfseries{$p$-value} \\
	\hline
	Exponential-Power law    & in  & -1133.5414 & 0.0031 \\
	Lognormal-Power law      & in  & 0.3770     & 0.8335 \\
	Lognormal-Exponential    & in  & 1133.9184  & 0.0030 \\
	DPLN-Power Law           & in  & -131.2078  & 2.1611e-20 \\
	DPLN-Lognormal           & in  & -131.5848  & 8.5745e-21 \\
	Exponential-Power law    & out & -1230.2352 & 0.0019 \\
	Lognormal-Power law      & out & -0.2428    & 0.1927 \\
	Lognormal-Exponential    & out & 1229.9925  & 0.0019 \\
	DPLN-Power Law           & out & -129.5294  & 2.2385e-21 \\
	DPLN-Lognormal           & out & -129.2866  & 3.4606e-21 \\
	\hline
\end{tabular}}
\label{tbl:rob_dd_comp}
\end{table}

\begin{table}[h]
\centering
\caption{Degree Distribution summaries}
\resizebox{0.48\textwidth}{!}{
\begin{tabular}{|c||c|c|c|c|}
	\hline

	\bfseries{Network} & 	\multicolumn{2}{c}{Zipf Distribution} &\multicolumn{2}{c|}{Lognormal Distribution}\\
	& \bfseries{In-deg. ($\alpha$)} & \bfseries{Out-deg ($\alpha$)} & \bfseries{In-deg ($mu$)} & \bfseries{In-deg ($\sigma$)}
	\\\hline\hline
	Humans & 3.1064 & 3.1195 & -9.5261 & 2.3997 \\
	\hline
	Robots & 2.1810 & 2.1810 & -0.9049 & 1.3495 \\
	\hline
\end{tabular}}
\label{tbl:dd_summary}
\end{table}

Depending on the exact distribution of the degree distributions of our networks, we can
draw conclusions about how they formed. Barab{\'a}si and Albert showed that
a random network model which has preferential attachment, where the probability of a
vertex sharing an edge with another vertex depends on their degrees, naturally leads to
a degree distribution following a power law~\cite{barabasi1999emergence}. Lognormal distributions
are generated by \emph{multiplicative} processes~\cite{limpert2001log}, wherein some variable $X_j$ at time $j$ is
determined by its previous state $X_{j-1}$ by multiplying it with some random variable $F_j$:
\begin{equation}
X_j = F_jX_{j-1}
\end{equation}
By taking the logarithm of both sides and applying the Central Limit Theorem, it can be shown that
$\log X_j$ follows a normal distribution as $j \to \infty$, and hence $X_j$ is asymptotically 
log-normally distributed. The presence of a log-normal degree distribution in our networks would indicate
that the average rate of increase in the degree of a node at time $t$ is proportionate to the degree at
that time. Thus, resources which are already popular become more popular faster than less commonly
requested resources. This ``rich get richer" behavior is common to processes underlying many
heavy-tailed distributions.

A process which produces a double-Pareto distribution related to DPLN is described by Mitzenmacher 
~\cite{mitzenmacher2004brief}. In this generalization of the multiplicative process, the time steps 
$T$ of the random variable $X_T$ are exponentially distributed. In particular, if the human network's
degree distribution is DPLN, this would imply that the times between the observation of requests
of new resources by humans is exponentially distributed. The observation of new resources in the
trace of human requests is reflected by the appearance of new nodes in the session graph, while
unique pairs of resources that appear in sequence within a session lead to the appearance of
new edges. The distribution of the creation times of new nodes and the dependence of
edge formation on the degrees of the nodes are what give rise to a DPLN distribution.

\section{Conclusion}
\label{sec:conclusion}
In this paper, we described a method for constructing a network called a 
\emph{session graph} from Web server logs based on the notion of a session described
in~\cite{tan2004discovery}. We looked at the basic properties of
these graphs for human and Web robot traffic on a university web server. 
We observed the presence of giant connected components using both weak and
strong connectivity, and studied qualitatively the rapid decrease in the
size of weakly connected components. We observed slight differences in
this decrease between human and robot networks, showing further that there
are differences in Web robot and human traffic.
We also carried out a comprehensive analysis of 
the degree distributions of the networks and find that they are 
best described 
by different theoretical distributions. This indicates important 
differences in the generative process for the respective networks. Of the distributions
considered, we found that the DPLN best describes the humans network,
while we were unable to distinguish between a power-law or log-normal
distribution for the robots network. We further found:
\begin{itemize}
\item that the densities of the human and robot session graphs are comparable;
\item that all session graphs have low reciprocity;
\item that a giant connected component (both weak and strong) is present in both
session graphs;
\item that the communities obtained by modularity maximization are more tightly 
      connected in the robots' session graph than the humans;
\item the degree distributions of both session graphs exhibit heavy-tailed
      behavior;
\item that a DPLN distribution best fits the
      degree distribution of the humans' session graph;
\item that the DPLN fit for the robots' session graph's degree distribution,
      but a log-normal or a Zeta distribution may be a reasonable fit for the humans'
      session graph. 
\end{itemize}

These findings lead to the following conclusions about
behavioral differences between human and Web robot traffic at \url{wright.edu}:
\begin{itemize}
\item{If a transition in one direction is observed, it is unlikely to be
      observed in the other direction. This may reflect the hyperlink structure
      of the website.}
\item{The existence of a giant connected component in the session graphs may 
      reflect the fact that the underlying web graph is almost fully connected, 
      i.e. starting from the homepage it is possible to reach almost every 
      resource on the Web server.}
\item{Robots may be more likely to transition between resources that are not
      connected by hyperlinks, as seen by the existence of fewer connected
      components and higher connectivity between communities in their
      session graph.}
\item{The time between the appearance of requests for resources that haven't
      been observed before may be exponential, under the assumption that
      the humans' session graph has a DPLN degree distribution.}
\item{Assuming heavy-tailed degree distributions, resources are more likely to 
	be observed in sequence if their degrees in the session graph are high.}
\end{itemize}

Future studies could
analyze data from multiple Web servers and compare their networks to identify
similarities and differences that arise from different types of Web traffic. For
example, by analyzing the degree distributions of networks constructed from traffic
from various Web servers, a study could be done to examine differences in these
distributions and produce hypotheses about the processes behind the network formation.
By understanding network formation, this tells us something about the characteristics
of Web robot and human traffic that could be used to improve prefetching and caching
algorithms. Another area for future work is in determining what constitutes a session.
For purposes of constructing a network representation of Web traffic, using a timeout may
not capture the properties of traffic as well as logical sessions using referers. An approach
that includes more than the time between requests could be used to improve session identification
and generate better network representations of the requests.
Finally, there were many network measures which were not considered. Centrality measures
such as eigenvector centrality, closeness centrality, and PageRank provide further distributions
to study for differences between human and Web robot session graphs. Other analyses that
could have been carried out are community detection and blockmodeling, which could be used
to find sets of resources which are somehow related. Future work could compute these
measures and partitions on smaller networks for tractability sake.

\section*{Acknowledgments}
We thank Logan Rickert for data processing support and Mark 
Anderson for providing the Wright State University server log data.
This paper is based on work supported by the National Science Foundation 
(NSF) under Grant No. 1464104. Any opinions, findings, and conclusions or 
recommendations expressed in this material are those of the author(s) and 
do not necessarily reflect the views of the NSF.

\bibliographystyle{IEEEtran}
\bibliography{icisdm2018graphs}

\begin{thebibliography}{10}
\providecommand{\url}[1]{#1}
\csname url@samestyle\endcsname
\providecommand{\newblock}{\relax}
\providecommand{\bibinfo}[2]{#2}
\providecommand{\BIBentrySTDinterwordspacing}{\spaceskip=0pt\relax}
\providecommand{\BIBentryALTinterwordstretchfactor}{4}
\providecommand{\BIBentryALTinterwordspacing}{\spaceskip=\fontdimen2\font plus
\BIBentryALTinterwordstretchfactor\fontdimen3\font minus
  \fontdimen4\font\relax}
\providecommand{\BIBforeignlanguage}[2]{{%
\expandafter\ifx\csname l@#1\endcsname\relax
\typeout{** WARNING: IEEEtran.bst: No hyphenation pattern has been}%
\typeout{** loaded for the language `#1'. Using the pattern for}%
\typeout{** the default language instead.}%
\else
\language=\csname l@#1\endcsname
\fi
#2}}
\providecommand{\BIBdecl}{\relax}
\BIBdecl

\bibitem{doran2011web}
D.~Doran and S.~S. Gokhale, ``Web robot detection techniques: overview and
  limitations,'' \emph{Data Mining and Knowledge Discovery}, vol.~22, no.~1,
  pp. 183--210, 2011.

\bibitem{doran2013comparison}
D.~Doran, K.~Morillo, and S.~S. Gokhale, ``A comparison of web robot and human
  requests,'' in \emph{Proceedings of the 2013 IEEE/ACM International
  Conference on Advances in Social Networks Analysis and Mining}.\hskip 1em
  plus 0.5em minus 0.4em\relax ACM, 2013, pp. 1374--1380.

\bibitem{chen2012challenges}
Y.-K. Chen, ``Challenges and opportunities of internet of things,'' in
  \emph{Design Automation Conference (ASP-DAC), 2012 17th Asia and South
  Pacific}.\hskip 1em plus 0.5em minus 0.4em\relax IEEE, 2012, pp. 383--388.

\bibitem{maier2010first}
G.~Maier, F.~Schneider, and A.~Feldmann, ``A first look at mobile hand-held
  device traffic,'' in \emph{International Conference on Passive and Active
  Network Measurement}.\hskip 1em plus 0.5em minus 0.4em\relax Springer, 2010,
  pp. 161--170.

\bibitem{broder2000graph}
A.~Broder, R.~Kumar, F.~Maghoul, P.~Raghavan, S.~Rajagopalan, R.~Stata,
  A.~Tomkins, and J.~Wiener, ``Graph structure in the web,'' \emph{Computer
  networks}, vol.~33, no.~1, pp. 309--320, 2000.

\bibitem{kleinberg1999web}
J.~M. Kleinberg, R.~Kumar, P.~Raghavan, S.~Rajagopalan, and A.~S. Tomkins,
  ``The web as a graph: measurements, models, and methods,'' in
  \emph{International Computing and Combinatorics Conference}.\hskip 1em plus
  0.5em minus 0.4em\relax Springer, 1999, pp. 1--17.

\bibitem{kumar2000web}
R.~Kumar, P.~Raghavan, S.~Rajagopalan, D.~Sivakumar, A.~Tompkins, and E.~Upfal,
  ``The web as a graph,'' in \emph{Proceedings of the nineteenth ACM
  SIGMOD-SIGACT-SIGART symposium on Principles of database systems}.\hskip 1em
  plus 0.5em minus 0.4em\relax ACM, 2000, pp. 1--10.

\bibitem{dikaiakos2005investigation}
M.~D. Dikaiakos, A.~Stassopoulou, and L.~Papageorgiou, ``An investigation of
  web crawler behavior: characterization and metrics,'' \emph{Computer
  Communications}, vol.~28, no.~8, pp. 880--897, 2005.

\bibitem{lee2009classification}
J.~Lee, S.~Cha, D.~Lee, and H.~Lee, ``Classification of web robots: An
  empirical study based on over one billion requests,'' \emph{computers \&
  security}, vol.~28, no.~8, pp. 795--802, 2009.

\bibitem{sisodia2015comparative}
D.~S. Sisodia, S.~Verma, and O.~P. Vyas, ``A comparative analysis of browsing
  behavior of human visitors and automatic software agents,'' \emph{American
  Journal of Systems and Software}, vol.~3, no.~2, pp. 31--35, 2015.

\bibitem{ihm2011towards}
S.~Ihm and V.~S. Pai, ``Towards understanding modern web traffic,'' in
  \emph{Proceedings of the 2011 ACM SIGCOMM conference on Internet measurement
  conference}.\hskip 1em plus 0.5em minus 0.4em\relax ACM, 2011, pp. 295--312.

\bibitem{donato2004large}
D.~Donato, L.~Laura, S.~Leonardi, and S.~Millozzi, ``Large scale properties of
  the webgraph,'' \emph{The European Physical Journal B-Condensed Matter and
  Complex Systems}, vol.~38, no.~2, pp. 239--243, 2004.

\bibitem{meusel2014graph}
R.~Meusel, S.~Vigna, O.~Lehmberg, and C.~Bizer, ``Graph structure in the
  web---revisited: a trick of the heavy tail,'' in \emph{Proceedings of the
  23rd international conference on World Wide Web}.\hskip 1em plus 0.5em minus
  0.4em\relax ACM, 2014, pp. 427--432.

\bibitem{sanders2015graph}
S.~Sanders and J.~Kaur, ``A graph theoretical analysis of the web using dns
  traffic traces,'' in \emph{Modeling, Analysis and Simulation of Computer and
  Telecommunication Systems (MASCOTS), 2015 IEEE 23rd International Symposium
  on}.\hskip 1em plus 0.5em minus 0.4em\relax IEEE, 2015, pp. 11--14.

\bibitem{liu2009user}
Y.~Liu, M.~Zhang, S.~Ma, and L.~Ru, ``User browsing graph: Structure, evolution
  and application.'' in \emph{WSDM (Late Breaking-Results)}, 2009.

\bibitem{page1999pagerank}
L.~Page, S.~Brin, R.~Motwani, and T.~Winograd, ``The pagerank citation ranking:
  Bringing order to the web.'' Stanford InfoLab, Tech. Rep., 1999.

\bibitem{calzarossa2011analysis}
M.~C. Calzarossa and L.~Massari, ``Analysis of web logs: challenges and
  findings,'' in \emph{Performance Evaluation of Computer and Communication
  Systems. Milestones and Future Challenges}.\hskip 1em plus 0.5em minus
  0.4em\relax Springer, 2011, pp. 227--239.

\bibitem{tan2004discovery}
P.-N. Tan and V.~Kumar, ``Discovery of web robot sessions based on their
  navigational patterns,'' in \emph{Intelligent Technologies for Information
  Analysis}.\hskip 1em plus 0.5em minus 0.4em\relax Springer, 2004, pp.
  193--222.

\bibitem{bvb_site}
\BIBentryALTinterwordspacing
(2017) Bots vs browsers. [Online]. Available:
  \url{http://www.botsvsbrowsers.com/}
\BIBentrySTDinterwordspacing

\bibitem{igraph_site}
\BIBentryALTinterwordspacing
(2017) igraph. The igraph Core Team. [Online]. Available:
  \url{http://igraph.org/}
\BIBentrySTDinterwordspacing

\bibitem{csardi2006igraph}
G.~Csardi and T.~Nepusz, ``The igraph software package for complex network
  research,'' \emph{InterJournal, Complex Systems}, vol. 1695, no.~5, pp. 1--9,
  2006.

\bibitem{newmannetworks}
M.~Newman, ``Networks: an introduction. 2010,'' \emph{United Slates: Oxford
  University Press Inc., New York}, pp. 1--2.

\bibitem{barabasi2013network}
A.-L. Barab{\'a}si, ``Network science,'' \emph{Philosophical Transactions of
  the Royal Society of London A: Mathematical, Physical and Engineering
  Sciences}, vol. 371, no. 1987, p. 20120375, 2013.

\bibitem{lewis2011network}
T.~G. Lewis, \emph{Network science: Theory and applications}.\hskip 1em plus
  0.5em minus 0.4em\relax John Wiley \& Sons, 2011.

\bibitem{fang2007new}
J.-q. Fang, X.-f. Wang, Z.-g. Zheng, Q.~Bi, Z.-r. Di, and L.~Xiang, ``New
  interdisciplinary science: Network science (1),'' \emph{PROGRESS IN
  PHYSICS-NANJING-}, vol.~27, no.~3, p. 239, 2007.

\bibitem{diaz2002survey}
J.~D{\'\i}az, J.~Petit, and M.~Serna, ``A survey of graph layout problems,''
  \emph{ACM Computing Surveys (CSUR)}, vol.~34, no.~3, pp. 313--356, 2002.

\bibitem{bastian2009gephi}
M.~Bastian, S.~Heymann, M.~Jacomy \emph{et~al.}, ``Gephi: an open source
  software for exploring and manipulating networks.'' \emph{ICWSM}, vol.~8, pp.
  361--362, 2009.

\bibitem{jacomy2014forceatlas2}
M.~Jacomy, T.~Venturini, S.~Heymann, and M.~Bastian, ``Forceatlas2, a
  continuous graph layout algorithm for handy network visualization designed
  for the gephi software,'' \emph{PloS one}, vol.~9, no.~6, p. e98679, 2014.

\bibitem{newman2006modularity}
M.~E. Newman, ``Modularity and community structure in networks,''
  \emph{Proceedings of the national academy of sciences}, vol. 103, no.~23, pp.
  8577--8582, 2006.

\bibitem{faloutsos1999power}
M.~Faloutsos, P.~Faloutsos, and C.~Faloutsos, ``On power-law relationships of
  the internet topology,'' in \emph{ACM SIGCOMM computer communication review},
  vol.~29, no.~4.\hskip 1em plus 0.5em minus 0.4em\relax ACM, 1999, pp.
  251--262.

\bibitem{clauset2009power}
A.~Clauset, C.~R. Shalizi, and M.~E. Newman, ``Power-law distributions in
  empirical data,'' \emph{SIAM review}, vol.~51, no.~4, pp. 661--703, 2009.

\bibitem{adamic2000power}
L.~A. Adamic and B.~A. Huberman, ``Power-law distribution of the world wide
  web,'' \emph{Science}, vol. 287, no. 5461, pp. 2115--2115, 2000.

\bibitem{foss2011introduction}
S.~Foss, D.~Korshunov, S.~Zachary \emph{et~al.}, \emph{An introduction to
  heavy-tailed and subexponential distributions}.\hskip 1em plus 0.5em minus
  0.4em\relax Springer, 2011, vol.~6.

\bibitem{reed2004double}
W.~J. Reed and M.~Jorgensen, ``The double pareto-lognormal distribution: a new
  parametric model for size distributions,'' \emph{Communications in
  Statistics-Theory and Methods}, vol.~33, no.~8, pp. 1733--1753, 2004.

\bibitem{zhang2015double}
C.~C. Zhang, ``The double pareto-lognormal distribution and its applications in
  actuarial science and finance,'' Master's thesis, Universit\'{e} de
  Montr\'{e}al, 2015.

\bibitem{alstott2014powerlaw}
J.~Alstott, E.~Bullmore, and D.~Plenz, ``powerlaw: a python package for
  analysis of heavy-tailed distributions,'' \emph{PloS one}, vol.~9, no.~1, p.
  e85777, 2014.

\bibitem{seshadri2008mobile}
M.~Seshadri, S.~Machiraju, A.~Sridharan, J.~Bolot, C.~Faloutsos, and
  J.~Leskove, ``Mobile call graphs: beyond power-law and lognormal
  distributions,'' in \emph{Proceedings of the 14th ACM SIGKDD international
  conference on Knowledge discovery and data mining}.\hskip 1em plus 0.5em
  minus 0.4em\relax ACM, 2008, pp. 596--604.

\bibitem{vuong1989likelihood}
Q.~H. Vuong, ``Likelihood ratio tests for model selection and non-nested
  hypotheses,'' \emph{Econometrica: Journal of the Econometric Society}, pp.
  307--333, 1989.

\bibitem{barabasi1999emergence}
A.-L. Barab{\'a}si and R.~Albert, ``Emergence of scaling in random networks,''
  \emph{science}, vol. 286, no. 5439, pp. 509--512, 1999.

\bibitem{limpert2001log}
E.~Limpert, W.~A. Stahel, and M.~Abbt, ``Log-normal distributions across the
  sciences: Keys and clues on the charms of statistics, and how mechanical
  models resembling gambling machines offer a link to a handy way to
  characterize log-normal distributions, which can provide deeper insight into
  variability and probability—normal or log-normal: That is the question,''
  \emph{BioScience}, vol.~51, no.~5, pp. 341--352, 2001.

\bibitem{mitzenmacher2004brief}
M.~Mitzenmacher, ``A brief history of generative models for power law and
  lognormal distributions,'' \emph{Internet mathematics}, vol.~1, no.~2, pp.
  226--251, 2004.

\end{thebibliography}

\end{document}